\documentclass[review]{elsarticle}

\usepackage{lineno,hyperref}
\modulolinenumbers[5]

\journal{Carbon}

\begin{document}

\begin{frontmatter}

\title{Valley Seebeck effect in gate tunable zigzag graphene nanoribbons}

\author[ad1,ad2]{Zhizhou Yu}
\author[ad1,ad3]{Fuming Xu\corref{cor}}
\cortext[cor]{Corresponding author}
\ead{xufuming@szu.edu.cn}
\author[ad1,ad2]{Jian Wang}
\address[ad1]{Department of Physics and the Center of Theoretical and Computational Physics, The University of Hong Kong, Pokfulam Road, Hong Kong, China}
\address[ad2]{The University of Hong Kong Shenzhen Institute of Research and Innovation, Shenzhen, China}
\address[ad3]{College of Physics Science and Technology, Shenzhen University, Shenzhen, 518060, China}

\begin{abstract}
We propose, for the first time, a valley Seebeck effect in gate tunable zigzag graphene nanoribbons as a result of the interplay between thermal gradient and valleytronics. A pure valley current is further generated by the thermal gradient as well as the external bias. In a broad temperature range, the pure valley current is found to be linearly dependent on the temperature gradient while it increases with the increasing temperature of one lead for a fixed thermal gradient. A valley field effect transistor (FET) driven by the temperature gradient is proposed that can turn on and off the pure valley current by gate voltage. The threshold gate voltage and on valley current are proportional to the temperature gradient. When the system switches on at positive gate voltage, the pure valley current is nearly independent of gate voltage. The valley transconductance is up to 30~$\mu$S if we take Ampere as the unit of the valley current. This valley FET may find potential application in future valleytronics and valley caloritronics.
\end{abstract}


\end{frontmatter}


\section{Introduction}
Thermoelectricity has been known since the observation of Seebeck effect in 1821 which revealed the interplay between thermal gradient and electric potential, opening a way for power generation and refrigeration\cite{Disalvo, Sales}. Recently, spin caloritronics, a new field combining thermoelectronics with spintronics that describes heat and spin transport, has attracted increasing attention\cite{Bauer1,Bauer2}. The spin Seebeck effect referring to the generation of spin voltage as a result of a temperature gradient has been observed experimentally in both ferromagnetic metals and magnetic insulators\cite{Uchida1,Uchida2,Jaworski}. Spin voltage, the potential between different spins, leads to a pure spin current which can be measured by the inverse spin Hall effect. On the other hand, the discovery of graphene\cite{Novoselov1,Novoselov2}, a two-dimensional atomically thin sheet of carbon atoms, opens a new path for the next generation green electronics\cite{Geim1,Geim2}. Zigzag graphene nanoribbons (ZGNRs), one-dimensional narrow stripes of graphene showing metallic characteristics, are particularly interesting for its potential applications in spintronics and thermoelectronics due to its unique electric properties and high thermal conductivity\cite{Tombros,Trauzettel,Hu}. The spin Seebeck effect has been proposed in ZGNR based materials from the first-principles calculation, which shows controllable thermal induced spin-polarized currents for graphene-based spin caloritronics\cite{Feng,Ni}.

Apart from the spin degree of freedom, graphene can also be characterized by its valley index, namely, the $K$ and $K'$ Dirac point in the Brillouin zone\cite{Geim1,Geim2}. According to the time reversal symmetry, the electron carrying one valley index shows the different direction of propagation from that of another valley\cite{Peres,Sarma,Abanin}. It is found that the intervalley coupling in suspended graphene is very weak\cite{McCann,Chen}. Therefore, being a good quantum number, the valley degree of freedom can be used in 'valleytronics' for the application of information processing similar to spin used in spintronics\cite{Abanin,Xiao,Nebel,Rycerz,White,Sun,Zettl,Chang}. A valley filter was firstly proposed in ZGNR based ballistic point contact in which the occupation of a single valley was achieved and such a valley polarization could be inverted by a local gate voltage\cite{Rycerz}. By introducing the line defect in graphene, a controllable 100\% valley polarization has been reported and widely studied theoretically\cite{White,Sun,Zettl}. Moreover, the generation of a pure bulk valley current without net charge current through quantum pumping induced by mechanical vibrations has been demonstrated in graphene by using the well-known Dirac Hamiltonian\cite{Chang}.

In this paper, we explore the valley degree of freedom in the Seebeck effect of ZGNRs, namely, how to generate a valley current by temperature gradient. Similar to the spin Seebeck effect, we call it valley Seebeck effect. A novel way of generating pure valley current is further proposed by applying both temperature gradient and bias voltage. We find that the pure valley current is linear with the thermal gradient in a broad temperature range while it increases significantly with the increasing temperature of leads under a same thermal gradient. We also present a gate tunable field effect transistor (FET) for the pure valley current driven by the temperature gradient, revealing a new perspective for valley caloritronics device applications.

\section{Model and formalism}
\begin{figure}
  \centering
  \includegraphics[width=3.4 in]{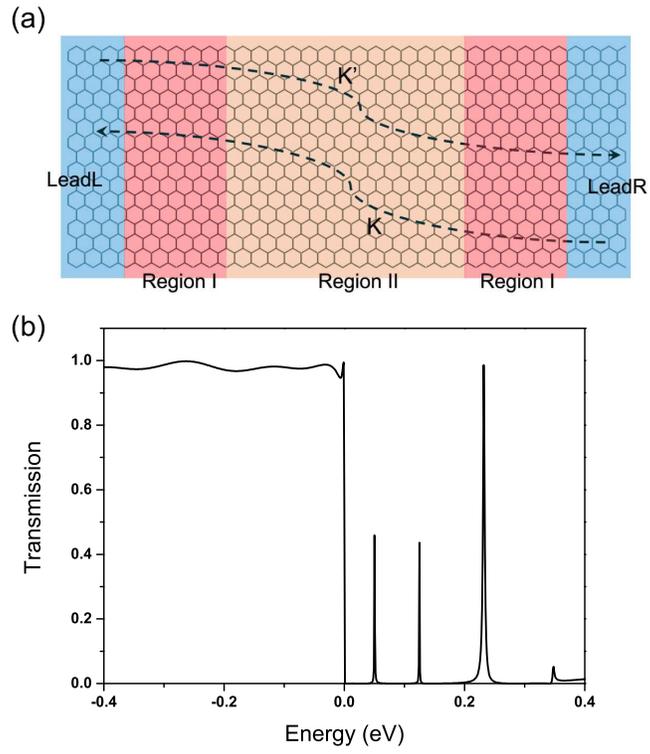}\\
  \caption{(a) Schematic diagram of ZGNRs with two semi-infinity leads (blue shadow), two static gate regions with $v_{g1} = 0.5$~V (red shadow) and $v_{g2}$ is tunable in the central region (orange shadow). (b) Transmission spectrum of ZGNRs with $v_{g1} = 0.5$~V and $v_{g2} = 0$. }
  \label{Fig1}
\end{figure}

The previous theoretical work shows that the lowest propagating mode for ZGNRs has a fixed valley index while for armchair nanoribbon each propagating channel is contributed by the mixed state of both valleys. As a result, the valley index or valley current for armchair nanoribbon is not well defined \cite{Rycerz}. Therefore, ZGNRs are chosen to study the valley effect. In order to control the valley current, two different gate regions are introduced into ZGNRs. The first gate with a constant voltage is applied in region \uppercase\expandafter{\romannumeral 1} and the second gate with tunable gate voltage is applied in region \uppercase\expandafter{\romannumeral 2} as indicated in Fig.~\ref{Fig1}(a). In the tight-binding approximation, the Hamiltonian for ZGNRs can be written as (here we set $\hbar = q = 1$ for simplicity),
\begin{equation}\label{ham}
  H = t\sum_{\langle i,j \rangle} c_i^\dag c_j + \mathrm{H.c.} + \sum_{m\in \mathrm{\uppercase\expandafter{\romannumeral 1}}} v_{g1} c_m c_m^\dag + \sum_{n\in \mathrm{\uppercase\expandafter{\romannumeral 2}}} v_{g2} c_n c_n^\dag,
\end{equation}
where $c_i$ ($c_i^\dag$) annihilates (creates) an electron on site $i$ of ZGNRs. $v_{g1}$,$v_{g2}$ denote the gate voltage introduced in region \uppercase\expandafter{\romannumeral 1} and region \uppercase\expandafter{\romannumeral 2}, respectively. $\langle ... \rangle$ refers to the nearest-neighboring sites and $t$ is the nearest neighbor hopping energy which is set to be 2.7~eV\cite{Geim2}. In our calculation, the width and length of the proposed ZGNRs based valley Seebeck device are 28.4~nm and 99.7~nm, respectively. The length of each gate region is set to be 25.8~nm.

The electric current can be obtained from the Landauer-B\"{u}ttiker formula,
\begin{equation}\label{cur}
  I = \int \frac{dE}{2\pi} (f_L - f_R) T(E),
\end{equation}
where $f$ is the Fermi-Dirac distribution
defined as,
\begin{equation}\label{fer}
f_\alpha(E,T) = \frac{1}{\exp [(E-E_F)/k_B T_\alpha] + 1},
\end{equation}
with the Fermi energy $E_F$, the Boltzmann constant $k_B$, and the temperature $T_\alpha$ in lead $\alpha$. $T(E)$ is the transmission coefficient,
\begin{equation}\label{tran}
T(E) = \mathrm{Tr} [\Gamma_L G^r \Gamma_R G^a ],
\end{equation}
where $G^{r(a)}$ is the retarded (advanced) Green's function and $\Gamma_\alpha$ is the linewidth function of lead $\alpha$. Denoting $I_K$ and $I_K'$ as the particle current of electron carrying valley index $K$ and $K'$, respectively, we define the valley current as,
\begin{equation}\label{valley}
  I_v = I_K -I_{K'}.
\end{equation}
From the band structure of ZGNRs, it can be found that the momentum and valley index of electron in the first subband are locked together so that the left-moving electron has valley index $K$ while electron with valley index $K'$ moves to the right as shown in Fig.~\ref{Fig1}(a). Such a unique electronic property of ZGNRs is independent of the ribbon width. Because at a given energy the sign of $f_L-f_R$ determines the direction of electron flow and hence valley index, we can express the valley current of ZGNRs as,
\begin{equation}\label{cur1}
  I_v = \int \frac{dE}{2\pi} \mathrm{sgn}(f_L - f_R) (f_L - f_R) T(E).
\end{equation}
which will be used in the calculation of valley current. To make the comparison with electric current, in the following calculation, we will take the unit of electric current as the unit of valley current.

\section{Results and Discussion}
It is well known that there is a conductance plateau of 1G$_0$ around the Fermi level for pristine ZGNRs which is symmetrical about the Fermi level, i.e., $T(E)=T(-E)$ with respect to the Fermi level.  We also know that the difference of the Fermi-Dirac functions $f_L(E) - f_R(E)=f_R(-E)-f_L(-E)$ is an odd function of energy measured at the Fermi level. As a consequence, there is no electric current due to the temperature gradient since the contribution of current below and above the Fermi level exactly cancel each other in Eq.(\ref{cur}), while a pure valley current can be generated from the definition of Eq.~(\ref{cur1}). This is the simplest but trivial way to achieve the valley Seebeck effect.

In order to achieve the valley Seebeck effect that can be efficiently controlled, one has to break the symmetry of $T(E)=T(-E)$. We first apply a static gate voltage in region I with $v_{g1} = 0.5$~V so that the system becomes a double barrier tunneling structure. The transmission spectrum of the system is plotted in Fig.~\ref{Fig1}(b). We see that the transmission plateau below the Fermi level still remains while its transmission coefficient oscillating around 0.99G$_0$ instead of 1G$_0$. On the other hand, only several resonant peaks exist above the Fermi level. Therefore, the symmetry of the transmission coefficients with respect to energy is broken so that the non-trivial valley Seebeck effect could be studied.

\begin{figure}
  \centering
  \includegraphics[width=2.9 in]{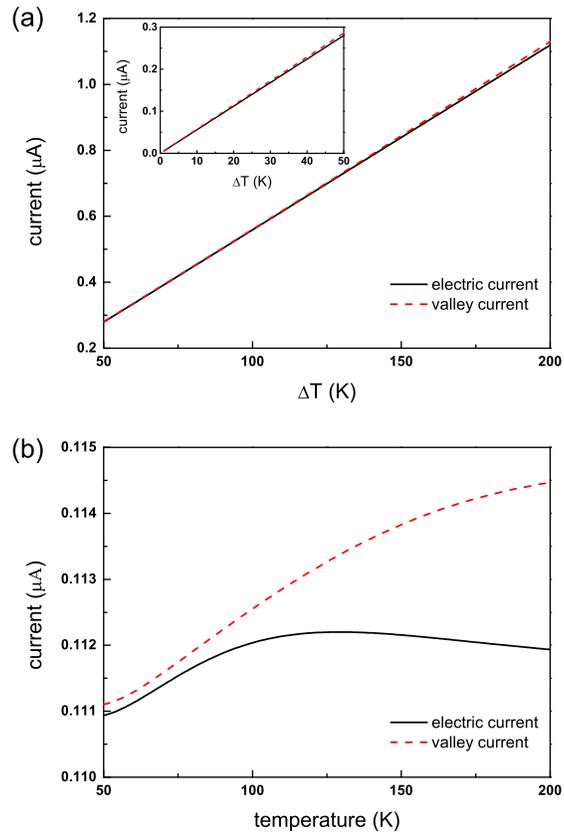}\\
  \caption{(a) Electric and valley currents as a function of $\Delta T$ with fixed $T_R=0$~K. Inset: Electric and valley currents as a function of $\Delta T$ with fixed $T_L=200$~K. (b) Electric and valley currents as a function of $T_L$ with fixed $\Delta T = 20$~K.}
  \label{Fig2}
\end{figure}

Figure~\ref{Fig2}(a) presents the electric and valley currents as a function of temperature gradient. The thermal induced electric current is caused by a temperature gradient ($\Delta T$) between the left electrode ($T_L$) and right electrode ($T_R$) without the external bias voltage. We first set the temperature of the right lead to be zero. We find that the electric current depends linearly on the temperature gradient and the differential thermoelectric conductance $dI/dT$ is 5.6~nA/K. The dependent of valley current on the temperature gradient is also linear with a slightly larger slope. The valley current at $\Delta T = 200$~K is only 10~nA higher than the electric current at the same temperature difference and the $dI_v/dT$ ratio is 5.7~nA/K. To further study the valley Seebeck effect in the linear regime, namely, $\Delta T < T$, we calculate the electric and valley current at different temperature gradients by fixing $T_L = 200$~K while keeping $T_L > T_R$ to generate the positive electric current, as shown in the inset of Fig.~\ref{Fig2}(a). Both the electric and valley current show linear dependence on the temperature gradient with the same differential thermoelectric conductance as those of $T_R = 0$~K, namely, 5.6~nA/K and 5.7~nA/K, respectively.

We also study the electric and valley current for a fixed temperature gradient while varying the temperatures of both leads. Fig.~\ref{Fig2}(b) shows the current as a function of the temperature of left lead with $\Delta T = 20$~K as an example. We find that the electric current first increases quickly with the increasing temperature and reaches a maximum about 112.2~nA at $T_L = 131$~K. It then decreases slightly due to the contribution of the transmission peak at 0.05 eV and the electric current reduces to 111.9~nA at $T_L = 200$~K. The valley current is higher than the electric current for all temperatures of left lead with the same fixed temperature gradient since all transmission coefficients contribute positively on it and it enhances significantly with the increasing temperature of left lead. The valley current is about 111.1~nA at $T_L = 50$~K and raises to 114.5~nA when the temperature of left lead increases to 200~K.

\begin{figure}
  \centering
  \includegraphics[width=3.2 in]{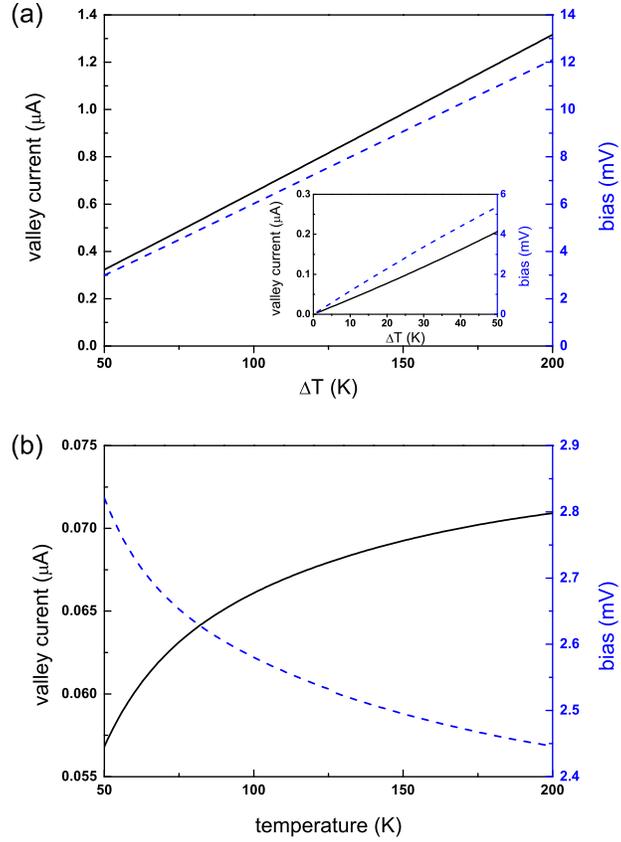}\\
  \caption{(a) Valley current (black solid line) and the corresponding external bias (blue dotted line) as a function of $\Delta T$ with fixed $T_R=0$~K. Inset: Valley current (black solid line) and the corresponding external bias (blue dotted line) as a function of $\Delta T$ with fixed $T_L=200$~K. (b) Valley current (black solid line) and the corresponding external bias (blue dotted line) as a function of  $T_L$ with fixed $\Delta T = 20$~K.}
  \label{Fig3}
\end{figure}

In order to obtain a pure valley current without the accompanying electric current, an external bias is then applied to balance the electric current. Fig.~\ref{Fig3}(a) presents the pure valley current and the applied bias as a function of temperature gradient with $T_R = 0$ K. An external bias is applied to make electric current vanish while the valley current is nonzero. Obviously, this bias depends on the temperature gradient linearly. We find that the bias increases from 3~mV at $\Delta T = 50$~K to 12.1~mV at $\Delta T = 200$~K with a $dV_b / dT$ ratio of 61~$\mu$V/K. In contrast to the vanished electric current, the valley current is enhanced slightly compared with the case without external bias as shown in Fig.~\ref{Fig2}(a) with a $dI_v / dT$ ratio of 6.6 nA/K. It is 1316~nA under the temperature gradient of 200~K, which is 186~nA higher than that with no external bias. The pure valley current and the corresponding external bias at different temperature gradients with $T_L = 200$~K are also plotted in the inset of Fig.~\ref{Fig3}(a). Both valley current and applied bias exhibit nearly linear characteristics with the $dI_v / dT$ and $dV_b / dT$ ratio of 4.1 nA/K and 107~$\mu$V/K, respectively. We find that the pure valley current is a little bit smaller than the valley current without external bias as shown in the inset of Fig.~\ref{Fig2}(a) at the same temperature gradient. For instance, at $\Delta T = 20$~K, the pure valley current is 77~nA with an external bias of 2.28~mV while the valley current without bias is 114~nA.

For a fixed temperature gradient of 20~K, the pure valley current and the corresponding applied bias as a function of the temperature of left lead are plotted in Fig.~\ref{Fig3}(b). We find that the valley current shows significant increase from 56.8~nA at $T_L = 50$~K to 70.9~nA at $T_L = 200$~K, while the corresponding external bias reduces from 2.82~mV to 2.45~mV because the contribution of the transmission coefficients above the Fermi level on the current becomes smaller at higher temperatures of both leads with the same temperature gradient.

\begin{figure}
  \centering
  \includegraphics[width=3 in]{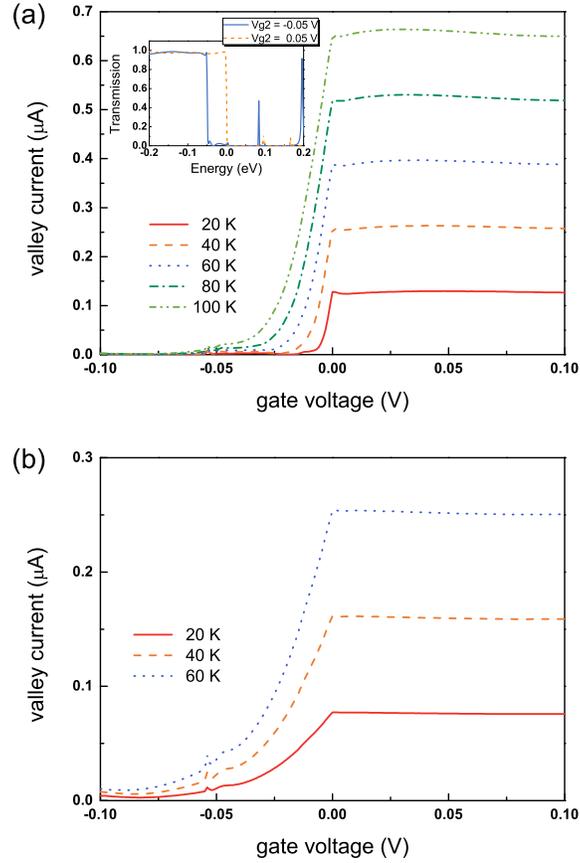}\\
  \caption{(a) Valley current as a function of $v_{g2}$ under several temperature gradients with fixed $T_R=0$~K. Inset: transmission spectrum of ZGNRs under $v_{g2} = \pm 0.05$~V with fixed $v_{g1} = 0.5$~V. (b) Valley current as a function of $v_{g2}$ under several temperature gradients with fixed $T_L=200$~K. }
  \label{Fig4}
\end{figure}

In order to control the pure valley current, we then introduce another gate voltage in region \uppercase\expandafter{\romannumeral 2} described in Fig.~\ref{Fig1}(a), namely, various values of $v_{g2}$ is used in the Hamiltonian as defined in Eq.~(\ref{ham}). This setup is a prototype of valley FET. The pure valley current as a function of gate voltage $v_{g2}$ under several temperature gradients with $T_R = 0$~K is plotted in Fig.~\ref{Fig4}(a). We find that under a negative gate voltage such as -0.05~V, the transmission plateau below the Fermi level shifts by 0.05~eV to the lower energy compared with that for the case of $v_{g2} = 0$ while the resonant transmission peaks above the Fermi level shift to the higher energy as shown in the inset of Fig.~\ref{Fig4}(a). Therefore, there is a transmission gap of 0.06~eV around the Fermi level leading to a very small valley current under a gate voltage of -0.05~V. When we increase the gate voltage, the transmission plateau moves towards the Fermi level which reduces the transmission gap, resulting in a significant increase of valley current. The threshold gate voltage and on valley current of the system are proportional to the temperature gradient and the valley current increases to a maximal value at the neutral gate voltage, which shows the potential application as a valley FET driven by the temperature gradient. The valley transconductances $dI_v/dv_{g2}$ of such a prototypical thermoelectrical valley FET are 27~$\mu$S for $\Delta T=20$~K and around 30~$\mu$S for the cases of $\Delta T=40, 60, 80$ and 100~K, which are almost twice the transconductance of typical silicon-based transistors\cite{silicon}. Upon further applying a positive gate voltage, the transmission plateau remains below the Fermi level rather than shifting to the higher energy. Therefore, the valley current is almost unchanged in the gate region of [0.0, 0.1] V, which leads to a stable operating valley current for the proposed ZGNR based thermoelectrical valley FET. Moreover, this stable valley current of ZGNRs under the positive gate voltage is also proportional to the temperature gradient.

We also calculate the pure valley current as a function of gate voltage $v_{g2}$ at different temperature gradients with $T_L = 200$~K. Similar to the case of $T_R = 0$~K, the valley current increases with the increasing gate voltage and remains at the maximum value when the gate voltage becomes positive. It is found that the on valley current of such a thermoelectrical valley FET is only about 70\% of that of the system with $T_R = 0$~K at the same temperature gradient, resulting in a smaller transconductance which is 7.5~$\mu$S for the case of $\Delta T =$60~K.

It is worth mentioning that the pure valley current studied in this paper may be detected experimentally from nonlocal multi-terminal measurements at room temperature\cite{Abanin}. The physical origin of pure valley current is due to the momentum-valley locking in the first subband of ZGNRs.  This property should be robust against a weak edge roughness since numerical results indicated that the Dirac cone of graphene is robust against disorders \cite{graphene}. We also wish to point out that in this study we have neglected the temperature redistribution according to the Joule heating originated from the external bias in our simulation which will affect the pure valley current in reality. Moreover, for other two-dimensional materials such as transition metal dichalcogenides (TMDC), the band structures of their nanoribbons are much more complicated and the phenomenon of momentum-valley locking no longer exists in TMDC. Hence our proposed theoretical method is not applicable to TMDC in the present stage.

\section{Conclusion}
In summary, we demonstrate the valley Seebeck effect in a gate tunable ZGNR based device by calculating valley current due to the temperature gradient across the device. For the first subband transport in ZGNRs, the pure valley current shows linear $I-T$ characteristics with the thermal gradient in a broad temperature range while it increases significantly with the increasing temperatures of the lead for a fixed temperature gradient. A thermoelectrical valley FET is then proposed. By tuning the gate voltage in the central region of ZGNRs, the valley FET can switch on and off with the valley transconductance up to 30~$\mu$S. It also shows a stable operating valley current under the positive gate voltage. Our theoretical work can be useful for the potential application of valley caloritronics.

\section*{Acknowledgments}

This work was financially supported by the Research Grant Council (Grant No. HKU 705212P), the University Grant Council (Contract No. AoE/P-04/08) of the Government of HKSAR and the National Natural Science Foundation of China (Grant No. 11374246). F.~Xu is supported by the National Natural Science Foundation of China (Grant No. 11447159).

\section*{References}
%

\end{document}